\documentclass[manuscript,screen]{acmart}
\AtBeginDocument{%
  }

\setcopyright{acmlicensed}
\copyrightyear{2026}
\acmYear{2026}
\setcopyright{cc}
\setcctype{by}
\acmConference[LAK 2026]{LAK26: 16th International Learning Analytics and Knowledge Conference}{April 27-May 01, 2026}{Bergen, Norway}
\acmBooktitle{LAK26: 16th International Learning Analytics and Knowledge Conference (LAK 2026), April 27-May 01, 2026, Bergen, Norway}
\acmPrice{}
\acmDOI{10.1145/3785022.3785093}
\acmISBN{979-8-4007-2066-6/2026/04}

\usepackage{subcaption}
\usepackage{threeparttable}
\usepackage{multirow} 
\usepackage{caption}
\usepackage[shortlabels]{enumitem}
\setlist[enumerate]{nosep}

\begin{document}


\title[Let Me Try Again]{Let Me Try Again: Examining Replay Behavior by Tracing Students' Latent Problem-Solving Pathways}




\author{Shan Zhang} 
\email{zhangshan@ufl.edu}
\orcid{0009-0003-3532-0661}
\affiliation{
  \institution{University of Florida}
  \city{Gainesville}
  \state{FL}
  \country{USA}
}

\author{Siddhartha Pradhan}
\email{sppradhan@wpi.edu}
\orcid{0009-0004-9977-1442}
\affiliation{
  \institution{Worcester Polytechnic Institute}
  \city{Worcester}
  \state{Massachusetts}
  \country{USA}
}

\author{Ji-Eun Lee}
\email{jieun_lee@sutd.edu.sg}
\orcid{0000-0001-8521-8997}
\affiliation{
  \institution{Singapore University of Technology and Design}
  \city{Singapore}
  \country{Singapore}
}

\author{Ashish Gurung}
\email{agurung@andrew.cmu.edu}
\orcid{0000-0001-7003-1476}
\affiliation{
  \institution{Carnegie Mellon University}
  \city{Pittsburgh}
  \state{Pennsylvania}
  \country{USA}
}

\author{Anthony F. Botelho}
\email{abotelho@coe.ufl.edu}
\orcid{0000-0002-7373-4959}
\affiliation{
  \institution{University of Florida}
  \city{Gainesville}
  \state{FL}
  \country{USA}
}

\renewcommand{\shortauthors}{Shan Zhang et al.}
\begin{abstract}
Prior research has shown that students' problem-solving pathways in game-based learning environments reflect their conceptual understanding, procedural knowledge, and flexibility. Replay behaviors, in particular, may indicate productive struggle or broader exploration, which in turn foster deeper learning. However, little is known about how these pathways unfold sequentially across problems or how the timing of replays and other problem-solving strategies relates to proximal and distal learning outcomes. This study addresses these gaps using Markov Chains and Hidden Markov Models (HMMs) on log data from 777 seventh graders playing the game-based learning platform of From Here to There!. Results show that within problem sequences, students often persisted in states or engaged in immediate replay after successful completions, while across problems, strong self-transitions indicated stable strategic pathways. Four latent states emerged from HMMs: Incomplete-dominant, Optimal-ending, Replay, and Mixed. Regression analyses revealed that engagement in replay-dominant and optimal-ending states predicted higher conceptual knowledge, flexibility, and performance compared with the Incomplete-dominant state. Immediate replay consistently supported learning outcomes, whereas delayed replay was weakly or negatively associated in relation to Non-Replay. These findings suggest that replay in digital learning is not uniformly beneficial but depends on timing, with immediate replay supporting flexibility and more productive exploration.

\end{abstract}




\maketitle

\section{Introduction} 
Learning from practice is not simply a matter of doing more. Research shows that the strategic intent and quality of practice may be just as, if not more important than, the sheer quantity of opportunities \cite{ericsson1993role, cepeda2006distributed, dunlosky2013improving}. In particular, studies have found that practice designed to deliberately vary problems and encourage learners to compare and explain alternative methods (e.g. through interleaving or side-by-side example comparisons) can support deeper knowledge as well as learning transfer (i.e. the ability to apply practiced skills in unfamiliar topics \cite{star2022exploring}) and improved recall \cite{alfieri2013learning, firth2021systematic, stanton2021fostering}; by purposefully engaging in different problem-solving strategies, students can become more attuned to key problem features and the affordances of different approaches, which enables them to move beyond the rote application of procedure toward a stronger conceptual understanding \cite{rohrer2015interleaved, alfieri2013learning, rittle2009compared}.

This adaptability, as fostered through practice, has often been studied under the construct of \textit{procedural flexibility} \cite{star2005reconceptualizing}, which can be operationalized as the capacity to generate multiple solution strategies for a given problem and the ability to adaptively select among them. Procedural flexibility reflects both the breadth and depth of a learner's skill and knowledge, highlighting not only whether students can produce alternative approaches but also whether they can recognize which strategies are most effective under particular conditions. One common way to cultivate this flexibility is by revisiting practice problems with alternative solution methods \cite{rittle2012developing}; such an approach allows for structured comparisons to discern connections between surface features and deeper principles across problems while also revealing when certain strategies are effective or prone to error. These processes are closely linked to the development of procedural fluency, deeper conceptual understanding, and greater knowledge transfer.

In game-based learning environments, opportunities to develop and demonstrate procedural flexibility emerge in distinctive ways. In such contexts, students rarely follow a single uniform path to a solution \cite{liu2022uncovering, pradhan2024gamification}. Instead, they navigate diverse problem-solving pathways (i.e. the distinct series of attempts students make within or even across learning opportunities as they progress or revisit earlier problems), ranging from exploratory trial-and-error to more intentional strategies that build on prior knowledge. Research has shown that these varied pathways not only capture differences in efficiency but also reflect important aspects of students’ conceptual understanding, procedural knowledge, and flexibility in mathematics learning \citep{liu2022uncovering, kai2018decision, Author2025}. Examining transitions among pathways is critical, as shifts can signal moments of struggle (both productive \cite{warshauer2015productive} and unproductive \cite{kai2018decision, fancsali2020towards}), disengagement \cite{gobert2015operationalizing}, or adaptive strategy use.

Building on these considerations, this paper examines how students navigate problem-solving pathways both within and across problems using data from \textit{From Here to There! (FH2T)} \cite{pradhan2024gamification}, a gamified version of \textit{Graspable Math} \cite{li2024math}; similar to other platforms such as \textit{Spatial Temporal (ST) Math} \cite{liu2017antecedents}, FH2T is designed to support procedural flexibility by offering learners the opportunity to re-attempt, or ``replay,'' problems as an optional feature rather than a structured requirement. Prior work in FH2T has categorized pathways as optimal, suboptimal, or incomplete, showing that exploratory attempts correlate with student learning, while frequent reliance on optimal solutions may reduce flexibility \cite{pradhan2024gamification}. It remains unclear, however, how sequences and transitions among pathways --- particularly around replayed attempts --- relate to learning and measures of student flexibility. The present study addresses these gaps by leveraging probabilistic sequence models to capture and analyze the dynamics of students' problem-solving pathways to examine how emergent strategies relate to both proximal (short-term) and distal (long-term) learning outcomes. Specifically, we seek to address the following research questions:
\begin{enumerate}
    \item[\textbf{RQ 1.}] How do students transition among different problem-solving strategies, both within individual problems and across their broader problem sequences?

    \item[\textbf{RQ 2.}] (a) What latent problem-solving states emerge when modeling students’ complete problem sequences? (b) What prominent transitions emerge across these states?
    
    \item[\textbf{RQ 3.}] To what extent do students’ latent problem-solving states predict their proximal and distal learning outcomes across-problem levels?

    \item[\textbf{RQ 4.}] How are students’ replay behaviors associated with their proximal and distal learning outcomes?

\end{enumerate}



\section{Related Work}  

\subsection{Sub-constructs of Algebraic Learning} 
Mastery over algebra requires not only conceptual and procedural knowledge, but also the ability to apply this knowledge \textit{flexibly} \cite{schneider_relations_2011}. \citeauthor{rittle-johnson_developing_2001} define conceptual understanding as the verbal and non-verbal knowledge of algebraic concepts and principles, such as algebraic symbols and conventions \cite{rittle-johnson_developing_2001}. In complement, procedural knowledge concerns the ordering of algebraic steps or transformations (i.e., the knowledge of the rules and procedures) required to solve a problem \cite{rittle-johnson_developing_2001}. These two forms of knowledge are closely related and often intersect; while conceptual knowledge enables the acquisition of procedural knowledge \cite{rittle-johnson_not_2015}, purposeful and sustained procedural practice can also build and strengthen conceptual knowledge \cite{rittle-johnson_developing_2001, schoenfeld_problem_2007}.

Together, conceptual and procedural knowledge provide the foundation for math flexibility, defined as a student's capacity to generate multiple potential solutions and then choose the one that is most efficient or appropriate for a given problem \cite{rittle2012developing}. Developing this flexibility not only improves problem-solving efficiency but also supports transfer to unfamiliar problems where reliance on a single memorized procedure may be insufficient.

Flexibility has been recognized internationally as a core goal of mathematics instruction \cite{hong_systematic_2023}, but fostering it in practice is challenging. Instruction often emphasizes a single ``preferred'' method for solving problems, with given practice opportunities tending to reinforce that approach. Prior works suggest that students benefit when they are encouraged to compare and contrast different strategies \cite{verschaffel_conceptualizing_2009, alfieri2013learning, firth2021systematic, stanton2021fostering}. Achieving this requires exposure to multiple approaches and a well-developed understanding of algebraic structures such as equivalence and grouping, which allow students to recognize when strategies are valid and applicable \cite{knuth_does_2006, ottmar_teaching_2012, welder_improving_2012}. However, few studies have empirically examined how the sequences of students' problem-solving behaviors, particularly during re-attempts or replays, impact the development of conceptual knowledge, procedural skill, and mathematical flexibility in algebra.

\subsection{Problem-solving Pathways in Game-based Learning} 
Students in game-based learning environments often employ different processes or follow different problem-solving pathways when approaching practice opportunities \cite{liu2022uncovering}. Some learners may begin with exploratory trial-and-error, gradually discerning the underlying rules or solutions through repeated attempts, whereas others adopt a more deliberate approach, planning strategically, weighing multiple solutions, before choosing how to best proceed. Understanding these pathways is important because they provide insight into students’ conceptual understanding, knowledge mastery, and skill development in learning games \cite{akcaoglu2021understanding, liu2022uncovering, spires2011problem}. For example, completing problems quickly and accurately using efficient strategies may be an indicator of conceptual understanding or knowledge mastery \cite{lee2025unpacking}, whereas over-reliance on supports or systematic guessing behavior may indicate struggle \cite{paquette2014towards, baker2008students}. Analyzing students' problem-solving pathways can therefore help teachers and instructional designers identify moments of productive or unproductive struggle \cite{warshauer2015productive, kai2018decision, fancsali2020towards} for the provision of targeted and timely feedback to support learning.

Within these pathways, one important form of engagement is whether students choose to replay problems when given such opportunities. Replay has been studied as a key behavior that emerges in game-based learning \cite{adetunji2024unlocking, gunter2008taking, harred2019long}. Replay behavior can be distinguished into two types: (1) \textit{pass attempt} as a method of restarting or re-trying a challenge, problem, or level before completing as a means of achieving a particular goal that they may have otherwise missed (e.g. achieving completion or in pursuit of a higher ``score,'' however it may be measured), and (2) \textit{elective replay}, in which students voluntarily choose to re-attempt a challenge, problem, or level after already completing a problem or passing a level as a means of trying new strategies or approaches that were previously not explored or potentially underutilized \cite{liu2022uncovering, zhang2022grade}. In this study, we use the term \textit{replay} to refer specifically to elective replay (i.e. type `2'), and \textit {reset} to describe attempts made to retry problems before any completion criterion is met (i.e. type `1').

Although relatively few studies have examined replay in detail, existing research indicates that it can be positively associated with performance and engagement \cite{adetunji2024unlocking}. Replay provides students with greater exposure to learning material and provides them with more control over their learning. However, the benefits of replay may depend on several contextual factors, including timing, the strategies implemented or explored, and other task demands \cite{harred2019long, liu2017antecedents, zhang2022grade}. For example, \citeauthor{liu2017antecedents} investigated elective replay among 5th graders in the game ST Math; they found that students who replayed the same level right after passing it achieved the highest performance, while those who struggled on a level tended to replay a different level during or after pass attempts \cite{liu2017antecedents}. These findings suggest that the impact of elective replay on performance varies with \textit{when} and \textit{how} replay occurs.

\section{Study Context: From Here to There! (FH2T)} 
\textit{From Here To There!} is a gamified learning application based on perceptual learning, embodied cognition, and gamification \cite{ottmar2015getting}. FH2T focuses on the structure of algebraic expressions to help improve students' knowledge, allowing students to think flexibly about mathematical operations \cite{chan2022slow}. Prior works have found that FH2T improves students' understanding of equivalence, mathematical flexibility, and conceptual and procedural knowledge \cite{chan2022slow, decker2023impacts, ottmar2015getting}.

\begin{figure}[ht]
    \centering
    \includegraphics[width=0.7\linewidth]{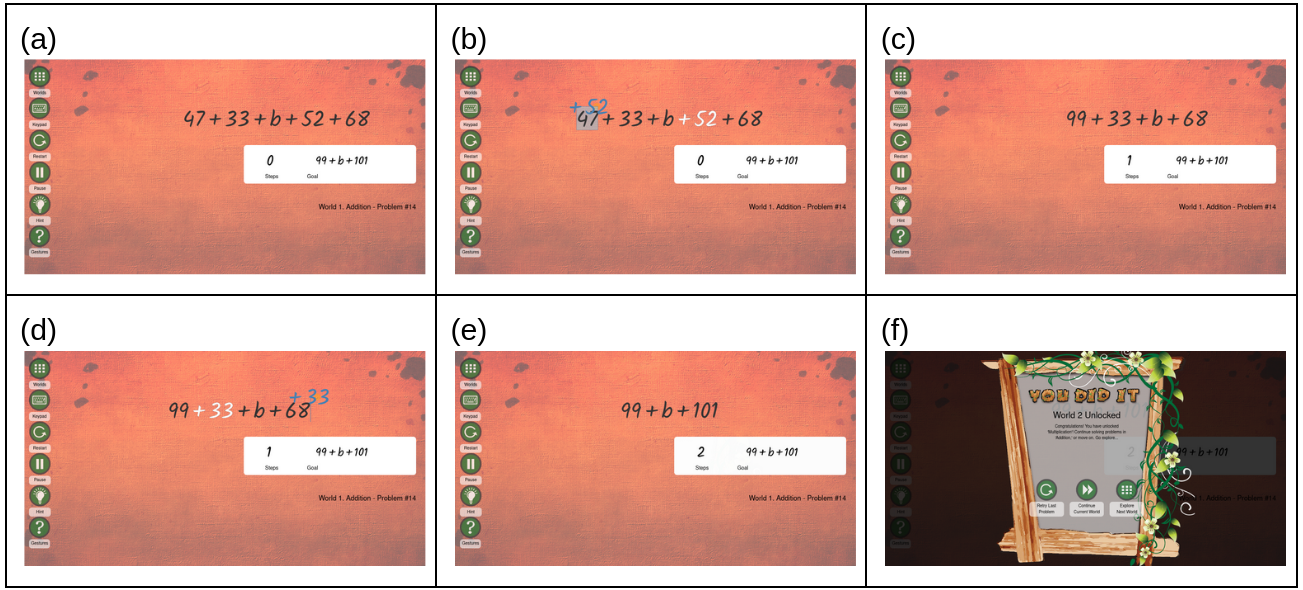}
    \caption{Example Problem 14 from FH2T with sample solution steps (Start State: $47+33+b+52+68$ Goal State: $99+b+101$}
    \label{fig:fh2t_example}
\end{figure}

FH2T presents algebraic equations through interactive virtual objects, which students manipulate using dynamic actions such as tapping or dragging. These interactions emphasize that algebraic transformations are processes in motion rather than static procedural steps \cite{ottmar2015getting}. Students progress by solving puzzle-like tasks: each problem begins with an initial expression (the start state) that must be transformed into a mathematically equivalent but perceptually different target expression (the goal state; see Fig.~\ref{fig:fh2t_example}). In that example, the expression ``47+33+b+52+68'' can be validly transformed into ``99+b+101'' through a sequence of algebraic moves. The start state can be transformed into an intermediate expression: ``99+33+b+68'' by adding ``47'' and ``52'' (see Fig~\ref{fig:fh2t_example}b \& ~\ref{fig:fh2t_example}c). Next, the problem can be solved by adding ``68'' and ``33'', resulting in the goal state: ``99+b+101'' (see Fig~\ref{fig:fh2t_example}d \& ~\ref{fig:fh2t_example}e). 

In total, the game consists of 252 problems that are grouped into 14 ``worlds'' based on mathematical concepts (e.g., multiplication, addition, fractions). Within each world, the problems are ordered based on difficulty; as such, students need to complete all non-optimal consecutive problems in one world to advance to the next world \cite{lee_does_2022}. After each completed problem attempt, the game rewards students based on the efficiency (i.e., number of steps) of their solution strategy (see Fig.\ref{fig:fh2t_example}).

Prior analyses of FH2T have examined how students navigate problems, with a focus on the distinct pathways they follow from start to goal. For instance, Pradhan et al.~\cite{pradhan2024gamification} used graph-based representations to identify three primary problem-solving pathways: optimal, suboptimal, and incomplete. These pathways are defined as follows:

\begin{itemize}
    \item \textbf{Optimal:} Pathways that solved the problem using the minimal number of algebraic moves. 
    \item \textbf{Suboptimal:} Pathways that reached the goal state but required more steps than the optimal solution. 
    \item \textbf{Incomplete:} Pathways that failed to reach the goal state.
\end{itemize}
The results from~\cite{pradhan2024gamification} showed that a higher usage of optimal and suboptimal pathways was correlated with higher post-test scores in comparison to Incomplete pathways. A follow-up study \cite{Author2025} further examined which types of pathways led to higher scores in different sub-constructs of algebraic learning: conceptual understanding, procedural knowledge, and math flexibility. Findings indicated that frequent optimal attempts were associated with lower flexibility scores. While the prior work examined how predictive aggregated measures of problem-solving pathways are correlated with algebraic knowledge \cite{pradhan2024gamification, Author2025}, the sequence and transitions of problem-solving pathways in relation to learning outcomes have not yet been investigated.

Prior research has found that replay in FH2T positively impacts learning outcomes \cite{lee_does_2022, vanacore_benefit_2023, liu2022reward}. For example, one study \cite{vanacore_benefit_2023} investigated whether 7th-grade students' propensity to replay was associated with their algebraic knowledge posttest scores. Results showed that each additional instance of replaying a problem was associated with higher post-test scores, suggesting that persistence behaviors may moderate the relationship between the gamified program and learning. However, it remains unclear whether all replay is beneficial for learning, regardless of when it occurs.

\section{Methods} 

\subsection{Dataset} 
To analyze students' transitions between different problem-solving strategies and pathways, we used data from a large Randomized Controlled Trial (RCT). This RCT was conducted during the COVID-19 pandemic between September 2020 and April 2021, and contains detailed action logs for the FH2T game. A total of 4,092 seventh-grade students were recruited from 11 middle schools in a major suburban district in the United States \cite{ottmar2023data}. The overall findings from the RCT have been reported in \cite{decker2023impacts}. The dataset is openly accessible to researchers via Open Science Framework (OSF)\footnote{This study was conducted as secondary data analysis under the guidance of a data sharing agreement. Information on how to access this data can be found here: \text{https://osf.io/r3nf2/}}


The assessment data for the RCT consist of pre-test and post-test scores taken before and after the intervention. Both these tests are comprised of 10 items (range 0-10), which measure conceptual knowledge (4 items), procedural knowledge (3 items), and math flexibility (3 items). These items in these tests are adapted from a previously validated measure and are scored based on correctness \cite{ottmar2023data}. Additionally, the assessment data contains students' state scores provided by the district for fifth and seventh grade (ranging from 265 to 740). The sixth-grade state test was not administered due to the COVID-19 pandemic \cite{ottmar2023data} and is unavailable in the dataset.  

In the RCT, 1,649 students were assigned to FH2T, of which 52.6\% were male and 47.4\% were female. Additionally, 49.8\% of the students identified as White, 24.8\% as Asian, and 16.4\% as Hispanic/Latino. The remaining 9\% identified with multiple racial categories \cite{decker2023impacts}. Of these students, 1,108 completed the pretest, and 777 of them completed the post-test. For the pretest, the mean values for the sub-constructs were: 1.69 for conceptual (\textit{SD} = 1.42), 1.57 for procedural (\textit{SD} = 0.98), and 1.46 for flexibility (\textit{SD} = 0.92). The overall mean for the pretest score was 4.71 (\textit{SD} = 2.68). Likewise, the mean post-test score was 4.50 (\textit{SD} = 2.93), with sub-constructs having mean values of: 1.89 for conceptual (\textit{SD} = 1.38), 1.42 for procedural (\textit{SD} = 1.04), and 1.19 for flexibility (\textit{SD} = 1.01). In addition, the dataset includes 5th-grade state assessment scores for 1,005 students (\textit{M} = 571.2, \textit{SD} = 64.0) and 7th-grade scores for 898 students (\textit{M} = 557.7, \textit{SD} = 64.4). In the FH2T game, students completed 111 problems (\textit{SD} = 55.05) on average out of a total of 252.


\subsection{Data Processing} 
We applied several preprocessing steps to remove noise and prepare the dataset for analysis. First, we removed student–problem rows that involved tutorial problems (\( \mathit{n} = 39 \)), as these were intended only for students to familiarize themselves with FH2T, and optional problems (\( \mathit{n} = 56 \)), as students are not required and can skip all these optional problems. Next, we filtered out problem–attempt rows with zero steps (\( \mathit{n} = 1{,}916 \)), with total \texttt{time\_spent} equal to 0 (\( \mathit{n} = 17{,}830 \)), and with \texttt{time\_spent} exceeding 1,800,000 milliseconds (30 minutes) (\( \mathit{n} = 29 \)). These cases likely reflect situations where students opened the platform but did not engage with the problem, or left the session running. Finally, we also removed students whose schools quit the study (\( \mathit{n} = 200 \)). After all cleaning and preprocessing, the final dataset contained 157,029 problem-solving attempts from 777 students across 157 unique problems.

To characterize students’ problem-solving trajectories, each problem-attempt row was assigned one of 12 pathway labels. An \texttt{optimal} attempt (2.3\%) denotes solving a problem with the most efficient sequence of steps, with \texttt{optimal\_end} (31.2\%) marking its final step. A \texttt{sub\_optimal} (4.4\%) attempt reaches a correct solution but requires more steps than the optimal path, with \texttt{sub\_optimal\_end} (11.1\%) indicating its conclusion. \texttt{Incomplete} attempts  (35.3\%)  capture partial or abandoned paths that do not reach the goal state, while \texttt{incomplete\_end}  (0.4\%)  marks their termination within a problem. Replay variants identify reattempts of previously completed problems: \texttt{replay\_optimal}  (0.7\%)  and \texttt{replay\_optimal\_end}  (4.8\%)  denote replayed optimal solutions, \texttt{replay\_sub\_optimal} (2.5\%) and \\ 
\texttt{replay\_sub\_optimal\_end}  (1.6\%) mark replayed suboptimal completions, and \texttt{replay\_incomplete} (5.3\%) and \\ \texttt{replay\_incomplete\_end}  (0.3\%)  correspond to replayed but unfinished attempts. Together, these labels differentiate efficient, exploratory, and disengaged behaviors across both initial and replayed attempts, enabling us to trace how strategies evolve within and across problems. A total of 157{,}029 labels were applied to form our analysis dataset.

To prepare the datasets for the \textit{within-problem} analysis, we sorted attempts by student~ID, problem~ID, and attempt number. This ensured that all attempts within a given problem were ordered together, allowing us to model students’ sequential behaviors within each problem without considering the broader timeline across problems. In contrast, for the \textit{across-problem} analysis, attempts were sorted strictly by student ID and timestamp, thereby reflecting the actual chronological order of students’ work across all problems. For example, a student could make three attempts on Problem~1, switch to Problem~2 for two attempts, and then return to Problem~1 for one additional attempt; this entire sequence would be captured by the \textit{across-problem} dataset. In contrast, in the \textit{within-problem} dataset, all four attempts on Problem~1 would be ordered together, representing the complete sequence of attempts for that problem. Thus, the \textit{across-problem} dataset preserved the full temporal sequence of problem-solving, whereas the within-problem dataset isolated sequences within individual problems independent of overall timing. 

\subsection{Analysis 1: Markov Chain Analysis of Problem-Solving Strategies Within and Across Problems}

To examine how students transitioned between different problem-solving strategies (\textbf{RQ 1}), we modeled their trajectories using first-order Markov chains. Each attempt was assigned one of the 12 labels introduced earlier (e.g., \texttt{incomplete}, \texttt{sub\_optimal}, \texttt{optimal}, and their corresponding \texttt{Replay} or \texttt{\_end} variants). We constructed two sets of transition matrices: (1) \textit{within-problem transitions}, where state sequences were restricted to attempts inside the same problem boundary, and (2) \textit{across-problem transitions}, where states were ordered by time across all problems attempted by a student

For both analyses, we encoded state labels into integers and counted all observed transitions between consecutive states. For each pair of states $(i,j)$, we computed (a) the transition count, (b) the transition probability $P(i \rightarrow j)$ obtained by row-normalization of counts, and (c) the average log-transformed time spent in state $i$ prior to moving to state $j$.  

The within-problem transition matrices capture micro-strategy dynamics inside single problems (e.g., ending a problem immediately after an incomplete attempt). The across-problem transition matrices capture students’ navigation and revisiting patterns over their entire problem-solving history (e.g., moving from a replayed suboptimal attempt on one problem to an incomplete attempt on the next). We visualized both sets of matrices using heatmaps: one for transition probabilities, and one for average log-time per transition. These representations highlight both the dominant state-to-state pathways and the temporal characteristics of students’ problem-solving strategies.

\subsection{Analysis 2: HMM Analysis of Problem-Solving Strategies Within and Across Problems}


To address \textbf{RQ 2}, we applied HMMs using the \texttt{hmmlearn} library in Python to identify latent problem-solving states and examine how students transition among these states at different levels of granularity. HMM was chosen because it is well-suited for modeling sequential data with unobserved structures, allowing us to infer underlying states from students’ observable actions \cite{rabiner2003introduction}. Each observation consisted of a categorical indicator of the attempt class (e.g., \texttt{optimal}, \texttt{sub\_optimal}, \texttt{incomplete}, and \texttt{replay} variants; \( \mathit{n} = 12 \)), which was one-hot encoded prior to model fitting. Unlike clustering (K-means or hierarchical clustering), which assigns each student to a single static group based on similarity, HMMs capture dynamic latent states that evolve over time. So one student can appear in multiple states across their problem-solving history, with the model estimating both the likelihood of behaviors within each state and the probabilities of transitioning between them. That said, HMM states should be interpreted as dynamic pathways to capture distinct patterns of strategy use and their transitions across attempts rather than fixed clusters.

For the \emph{within-problem} analysis, we defined each session as a unique student--problem pair and sorted attempts by student, problem, and attempt order. This analysis emphasizes the micro-level dynamics of students’ strategy choices within individual problems. For the \emph{across-problem} analysis, we instead defined sessions at the student level and sorted all attempts chronologically across problems, which captures how strategies evolve over time.

\begin{figure}[b]
    \centering
    \begin{subfigure}{0.46\textwidth}
        \centering
        \includegraphics[width=\linewidth]{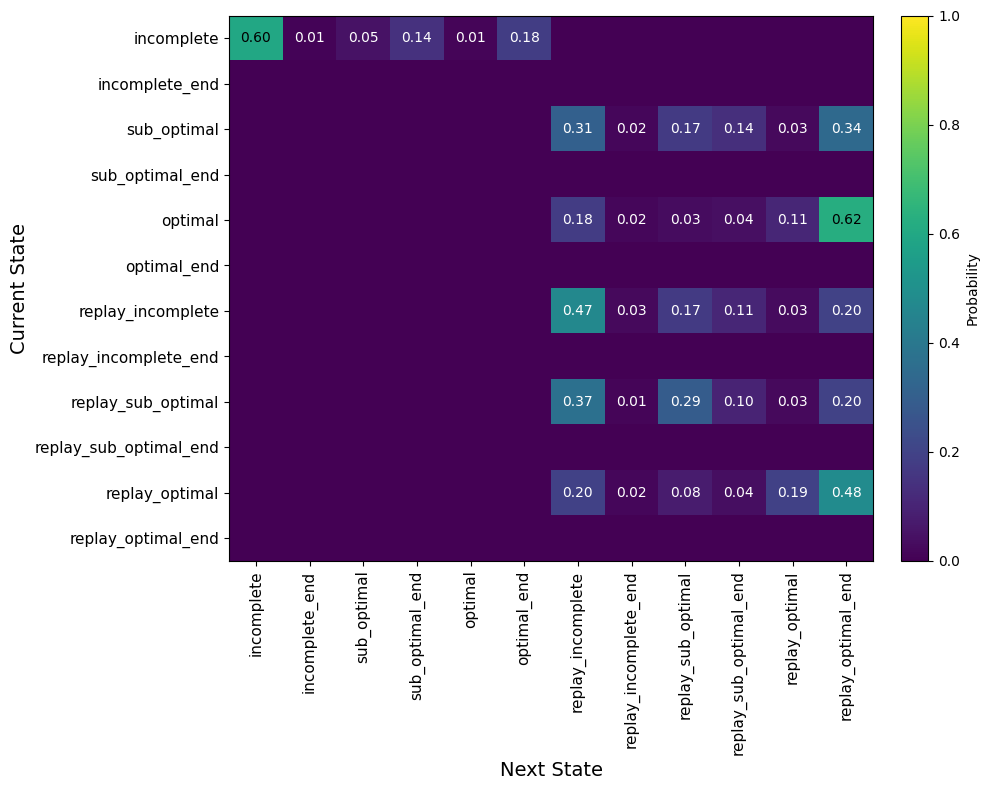}
        \caption{Within-problem Transition Probabilities}
        \label{fig:W_P}
    \end{subfigure}
    \hfill
    \begin{subfigure}{0.46\textwidth}
        \centering
        \includegraphics[width=\linewidth]{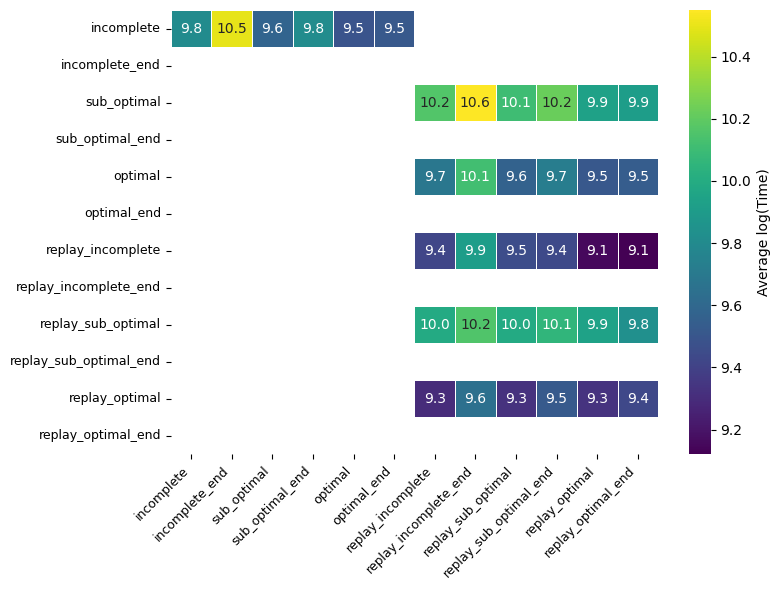}
        \caption{Within-problem Avg log(Time)}
        \label{fig:W_T}
    \end{subfigure}
    
    \vspace{0.1cm}
    
    \begin{subfigure}{0.46\textwidth}
        \centering
        \includegraphics[width=\linewidth]{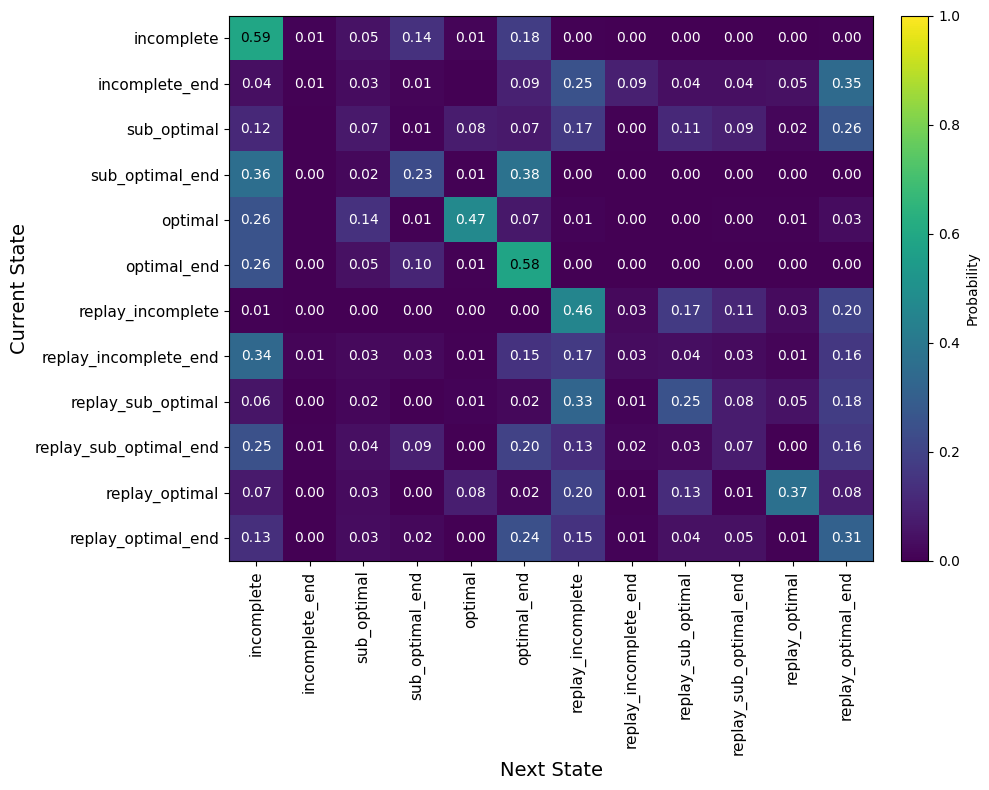}
        \caption{Across-problem Transition Probabilities}
        \label{fig:A_P}
    \end{subfigure}
    \hfill
    \begin{subfigure}{0.46\textwidth}
        \centering
        \includegraphics[width=\linewidth]{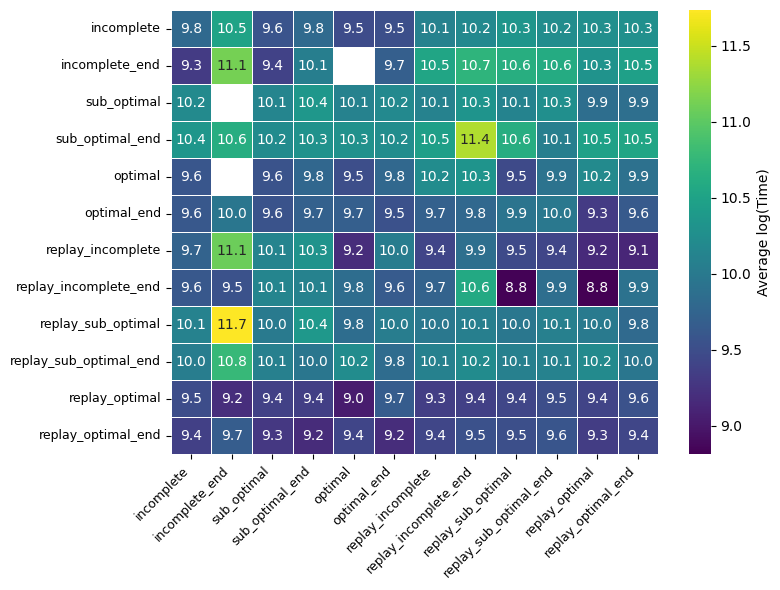}
        \caption{Across-problem Avg log(Time)}
        \label{fig:A_T}
    \end{subfigure}
    
    \caption{Transition matrices for within-problem (top row) and across-problem (bottom row) Markov Chain analyses. Each row shows transition probabilities (P) and average log(Time) (T).}
    \label{fig:transitions_PT}
\end{figure}

The number of hidden states was determined using 5-fold cross-validation with the Bayesian Information Criterion (BIC), where lower values indicate better model fit. After fitting the models, we decoded hidden state assignments for all attempts and extracted two outputs: (i) the emission distributions, which reveal the attempt classes most strongly associated with each hidden state; and (i) the transition probability matrix ($P(s_{t+1}\mid s_t)$), which characterizes how likely students are to move from one latent state to another. In addition, we also calculated empirical run-lengths, defined as the average number of consecutive steps assigned to each hidden state before transitioning, as well as student counts, average log time, average number of problems, and total problem counts associated with each state to better understand their role in students’ problem-solving trajectories.


\subsection{Analysis 3: Regression Analysis of Latent Problem-Solving Strategy States and Replays}

Regarding \textbf{RQ 3}, we examined how students’ proportional engagement in latent HMM states predicted their subsequent learning outcomes at both proximal and distal levels. For both the within- and across-problem HMM analyses, we first computed student-level summaries of hidden state engagement. For each student, we calculated the proportion of total attempts assigned to each hidden state (\emph{state percentages}) as well as the number of distinct problems attempted (\emph{problem count}). These measures provided student-level indicators of how strongly each learner’s trajectory was characterized by particular latent strategies and the breadth of their problem-solving experience.

We then merged these student-level summaries with assessment datasets. Proximal learning outcomes were measured using pre- and post-test scores at the total-math level as well as sub-constructs, Conceptual, Procedural, and Flexibility. Distal learning outcomes were captured using students’ standardized state test scores administered longitudinally, with Grade~5 serving as the pre-test (5th Grade State Test Score) and Grade~7 serving as the post-test (7th Grade State Test Score). In addition, log-derived features, total hints requested, were included to control for overall engagement. To enable interpretability, one hidden state was treated as the reference category (i.e., omitted from the regression), and the remaining states were entered as predictors.

We conducted Ordinary least squares (OLS) regression models using the \texttt{statsmodels} package in Python. For proximal outcomes, dependent variables were post-test scores (total and sub-constructs), with predictors including corresponding pre-test scores, HMM-derived state percentages, problem count, and total hints. For distal outcomes, Grade~7 state test scores were regressed on the same predictors, with Grade~5 scores included as a covariate.

In addition, for \textbf{RQ 4}, we distinguished between two types of replay behavior: students who \emph{immediately} retried the same problem in consecutive attempts (\emph{Immediate Replay}) and students who \emph{returned} to a problem after working on other problems (\emph{Delayed Replay}). To clarify these effects, specifically, we decomposed replay into three mutually exclusive categories: \emph{Immediate Replay} (consecutive attempts on the same problem), \emph{Delayed Replay} (a return to the same problem after intervening attempts on other problems), and \emph{Non-Replay} (any sequence that does not involve replay; reference group). We then calculated each student’s proportion of attempts in these categories and fit regression models using the same covariates as before. For all regressions, we applied the Benjamini--Hochberg procedure to control the false discovery rate (FDR) when adjusting the resulting $p$-values for multiple comparisons.

\section{Results} 

\subsection{Markov Chain Results}

Figure~\ref{fig:W_P} (within) and Figure~\ref{fig:A_P} (across) show that transition probabilities are sparse but concentrated on a few pathways. \textbf{Within problems}, the strongest transitions reflected persistence in ongoing states (e.g., \texttt{incomplete} $\rightarrow$ \texttt{incomplete}) and immediate replay after successful solutions (\texttt{optimal} $\rightarrow$ \texttt{replay\_optimal\_end}). Replay dynamics also emerged, with students often looping within replay categories before reaching a terminating ``end'' state.

\textbf{Across problems}, persistence remained the dominant pattern, with students often repeating the same strategy type from one problem to the next (e.g., \texttt{optimal} $\rightarrow$ \texttt{optimal}; \texttt{replay\_incomplete} $\rightarrow$ \texttt{replay\_incomplete}). Cross-category shifts were less frequent but notable. For instance, incomplete solutions often transitioned to terminal states on subsequent problems, and suboptimal endings frequently preceded optimal completions.


Figures~\ref{fig:W_T} and \ref{fig:A_T} further show that transition times were longer for suboptimal and replayed strategies, particularly when leading to termination states, whereas replay-to-replay endings tended to occur more quickly.

\subsection{Hidden Markov Model Results}
\begin{figure}[!h]
    \centering
    \begin{subfigure}{0.55\textwidth}
        \centering
        \includegraphics[width=\linewidth]{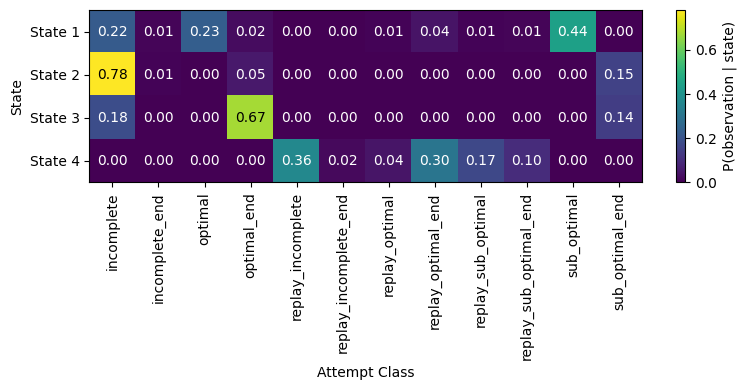}
        \caption{Across-problem HMM Emission probabilities}
        \label{fig:A_E_HMM}
    \end{subfigure}
    \hfill
    \begin{subfigure}{0.35\textwidth}
        \centering
        \includegraphics[width=\linewidth]{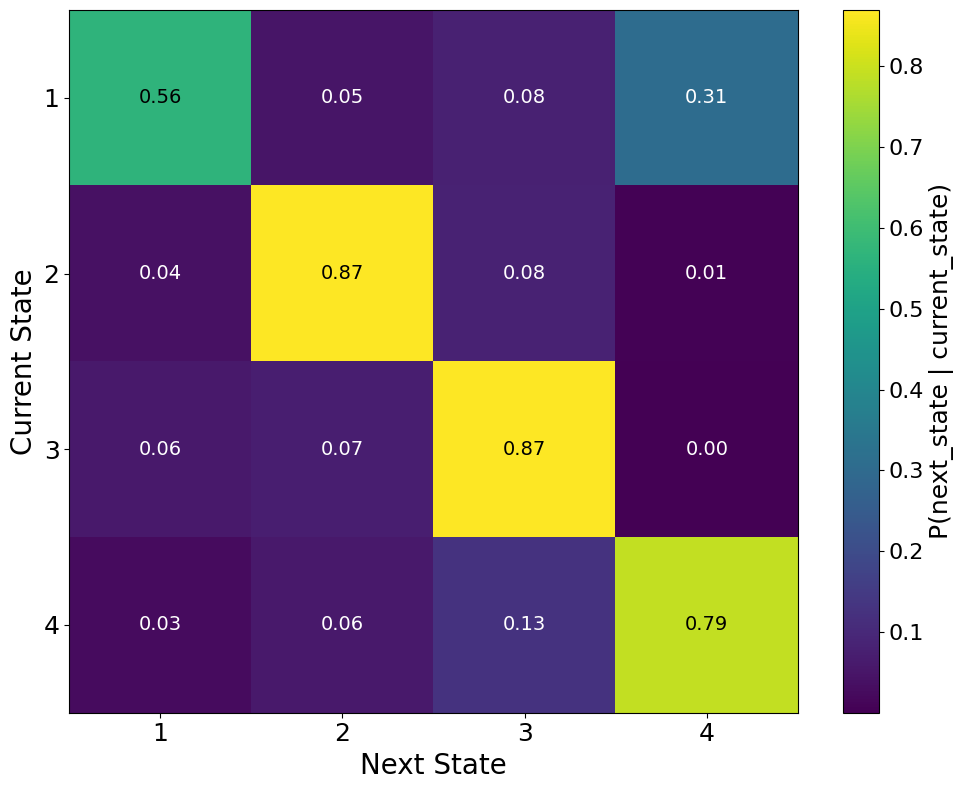}
        \caption{Across-problem HMM Transition Prob.}
        \label{fig:A_T_HMM}
    \end{subfigure}

    \caption{HMM results for Across-problem analysis: (a) emission distributions, and (b) transition probabilities.}
    \label{fig:across_HMM}
\end{figure}

We applied HMMs with 5-fold cross-validation and BIC-based model selection, resulting in a four-state solution (Table~\ref{tab:hmm_states}). The states were interpreted as follows (from Figure~\ref{fig:A_E_HMM}):  
\begin{itemize}

    \item \textbf{State~1 (Mixed):} A blend of incomplete, optimal, and suboptimal attempts, with occasional replay.  
    
    \item \textbf{State~2 (Incomplete-dominant):} Characterized by high probabilities of incomplete solutions.  
    
    \item \textbf{State~3 (Optimal-ending):} Dominated by problem completions through optimal or suboptimal endings.  
    
    \item \textbf{State~4 (Replay-dominant):} Defined by repeated replay behaviors across attempts. 
    
\end{itemize}  

The transition matrix (Figure~\ref{fig:A_T_HMM}) showed strong persistence in the incomplete-dominant (0.87), optimal-ending (0.87), and replay-dominant states (0.79). By contrast, the mixed state was less stable (0.56), frequently transitioning into replay cycles (0.31) or to optimal endings (0.08). Movement out of the incomplete-dominant state was rare but most often directed toward optimal endings, while the replay-dominant state sometimes transitioned into both optimal-ending and incomplete-dominant states.  

Run-length analyses (Table~\ref{tab:hmm_states}) reinforced these patterns. Students typically spent long stretches in incomplete-dominant (mean = 10.3 steps) and optimal-ending states (mean = 9.5 steps), moderate time in replay-dominant states (mean = 4.4), and only brief periods in the mixed state (mean = 2.3).

\begin{table}[t]
\centering
\caption{Summary of HMM States: Student Counts, Average Time, Problem Engagement, and Empirical Run-Lengths}
\label{tab:hmm_states}
\begin{tabular}{lccccc}
\hline
\textbf{HMM State} & \textbf{N Students} & \textbf{Avg. Time (Log)} & \textbf{Avg. Problems} & \textbf{Total Problems} & \textbf{Avg. Run-Length} \\
\hline
State 1 & 941  & 9.85 & 9.4  & 154 & 2.30 \\
State 2 & 1059 & 9.89 & 11.99 & 154 & 10.30 \\
State 3 & 1168 & 9.66 & 48.07 & 157 & 9.51 \\
State 4 & 933  & 9.54 & 8.6  & 153 & 4.41 \\
\hline
\end{tabular}
\end{table}

\subsection{Examining Relationships with Proximal and Distal Learning Outcomes}

We then examined whether the proportion of engagement students devoted to each HMM state predicts their learning outcomes. All results reported here were adjusted using the Benjamini–Hochberg correction for multiple comparisons (Tables~\ref{tab:regression_results_across_sub} and \ref{tab:regression_results_across}). Using State 2 (Incomplete dominant) as the reference group, several clear patterns emerged.

Students who spent more time in State 4 (Replay-dominant) showed positive associations with multiple outcomes, predicting higher conceptual ($b=1.20$, $p<.05$) and flexibility scores ($b=1.10$, $p<.05$), as well as stronger performance on the state test ($b=0.63$, $p<.01$), compared to the Incomplete group. Although post-math scores were also higher for State 4 ($b=1.87$), this effect did not reach statistical significance ($p=.066$). In addition, students who spent more time in State 3 (Optimal-ending) demonstrated higher conceptual scores ($b=0.61$, $p<.05$) and performed better on the state test ($b=0.34$, $p<.01$) in relation to the Incomplete group. A significant effect was also observed for flexibility ($p=0.041$); however, after applying the $p$-value correction, this result was no longer significant ($p=0.057$). Meanwhile, spending more time in State 1 (a mix of incomplete, suboptimal, and optimal behaviors) was not significantly related to any of the outcome measures.

Alongside these state-based findings, several behavioral covariates also emerged as reliable predictors. Students who solved more problems tended to perform better across all outcomes, while those who requested more hints scored lower on conceptual and flexibility measures (with only a marginal negative effect for procedural knowledge, $p=0.047$ before $p$-value correction). As expected, prior knowledge (pre-conceptual, pre-procedural, pre-flexibility, and earlier state test scores) was a strong predictor of corresponding post-measures.

Overall, these results suggest that students who engaged in pathways characterized by State 3 (Optimal-ending) and State 4 (Replay-dominant) benefited in terms of immediate conceptual understanding, flexibility (for State 4 only), and longer-term performance on standardized tests.

\begin{table*}[t]
\centering
\caption{HMM regression for the across-problem analysis on conceptual, procedural, and flexibility scores}
\label{tab:regression_results_across_sub}
\renewcommand{\arraystretch}{1.05}
\setlength{\tabcolsep}{8pt}
\small

\begin{threeparttable}
  \begin{tabular}{lll|ll|ll}
    \hline
    & \multicolumn{2}{c|}{\textbf{Conceptual}} 
    & \multicolumn{2}{c|}{\textbf{Procedural}} 
    & \multicolumn{2}{c}{\textbf{Flexibility}} \\
    \cline{2-7}
    \textbf{Predictor} 
      & $\boldsymbol{b}$ & \textbf{SE} 
      & $\boldsymbol{b}$ & \textbf{SE} 
      & $\boldsymbol{b}$ & \textbf{SE} \\
    \hline
    Intercept                           & -0.22 & 0.2   & 0.04 & 0.18  & -0.32 & 0.17 \\
    State 1 Proportion (Mix of Incomplete, Suboptimal, Optimal)      & -0.39 & 0.78   & -0.06 & 0.7   & -0.7 & 0.65 \\
    State 3 Proportion (Optimal-ending dominant) & 0.61* & 0.26   & 0.21 & 0.23   & 0.43$\dagger$  & 0.21 \\
    State 4 Proportion (Replay-dominant) & 1.2* & 0.51   & 0.8 & 0.45   & 1.1* & 0.43 \\
    Total hints requested               & -0.01* & 0.00   & -0.00 & 0.00   & -0.01*** & 0.00 \\
    Problem count                       & 0.01*** & 0.001   & 0.01** & 0.00   & 0.01*** & 0.00 \\
    Pre-conceptual Score                      & 0.4*** & 0.03   & -- & --   & -- & -- \\
    Pre-procedural Score                      &--  & --   & 0.27*** & 0.04   & -- & -- \\
    Pre-flexibility Score                      &--  & --   & -- & --   & 0.26*** & 0.04 \\
    \hline
    \textbf{Model fit}  
      & $\boldsymbol{R^2}$ & \textbf{N} 
      & $\boldsymbol{R^2}$ & \textbf{N} 
      & $\boldsymbol{R^2}$ & \textbf{N} \\
      & 0.47 & 777   & 0.26 & 777   & 0.31 & 777 \\           
    \hline
  \end{tabular}
  \begin{tablenotes}[flushleft]
    \footnotesize
    \item \textit{Note.} * $p<.05$, ** $p<.01$, *** $p<.001$. All $p$-values shown are Benjamini–Hochberg FDR–adjusted. $\dagger$ indicates results that were significant before adjustment but not after. Reference group: State 2 Proportion (Incomplete-dominant).
  \end{tablenotes}
\end{threeparttable}

\end{table*}

\begin{table*}[t]
\centering
\caption{HMM regression for the across-problem analysis on total math and state test scores}
\label{tab:regression_results_across}
\renewcommand{\arraystretch}{1.05}
\setlength{\tabcolsep}{10pt}
\small

\begin{threeparttable}
  \begin{tabular}{lll|ll}
    \hline
    & \multicolumn{2}{c|}{\textbf{Post-math Score}}
    & \multicolumn{2}{c}{\textbf{7th Grade State Test Score}} \\
    \cline{2-5}
    \textbf{Predictor} & $\boldsymbol{b}$ & \textbf{SE}
                       & $\boldsymbol{b}$ & \textbf{SE} \\
    \hline
    Intercept                           &-0.57    & 0.39 &-0.56***  &0.11   \\
    State 1 Proportion (Mix of Incomplete, Suboptimal, Optimal)     &-1.07 & 1.54  & -0.3  & 0.34  \\
    State 3 Proportion (Optimal-ending dominant) &0.66 &0.5  &0.34**   & 0.13 \\
    State 4 Proportion (Replay-dominant) &1.87  & 1.01 &0.63**  &0.24  \\
    Total hints requested               & -0.01$\dagger$  &0.00 &-0.00** &0.00 \\
    Problem count                       &0.03***  &0.00   &0.01*** &0.00 \\
    Pre-math Score                      &0.54***  &0.03  & --      & --   \\
    5th Grade State Test Score                         & --      & --    & 0.72*** & 0.02 \\
    \hline
    \textbf{Model fit}  & $\boldsymbol{R^2}$ & \textbf{N}
                       & $\boldsymbol{R^2}$ & \textbf{N} \\

                                          & 0.55   & 777  & 0.74   &  765 \\           
    \hline
  \end{tabular}
  \begin{tablenotes}[flushleft]
    \footnotesize
    \item \textit{Note.} * $p<.05$, ** $p<.01$, *** $p<.001$. All $p$-values shown are Benjamini–Hochberg FDR–adjusted. $\dagger$ indicates results that were significant before adjustment but not after. Reference group: State 2 Proportion (Incomplete-dominant).
  \end{tablenotes}
\end{threeparttable}

\end{table*}

\subsection{Investigating Students' Replay Behaviors in Relation to Proximal and Distal Learning Outcomes}


According to Tables~\ref{tab:regression_results_across_sub_replay} and \ref{tab:regression_results_replay}, students who engaged more frequently in Immediate Replay significantly outperformed the Non-Replay group across all outcome domains. Specifically, higher proportions of Immediate Replay were associated with stronger conceptual ($b = 1.39$, $SE = 0.46$, $p < .01$), procedural ($b = 1.47$, $SE = 0.41$, $p < .001$), and flexibility scores ($b = 1.54$, $SE = 0.38$, $p < .001$), as well as higher overall post-math scores ($b = 3.22$, $SE = 0.90$, $p < .001$) and state test performance ($b = 0.65$, $SE = 0.22$, $p < .01$). By contrast, greater proportions of Delayed Replay were not beneficial and, in some cases, predicted lower achievement in relation to Non-Replay. Specifically, Delayed Replay was negatively associated with flexibility ($b = -1.10$, $SE = 0.43$, $p < .01$), post-math scores ($b = -2.32$, $SE = 1.02$, $p < .05$), and state test performance ($b = -0.62$, $SE = 0.29$, $p < .05$).

Similar to prior models, covariates showed consistent effects: students who solved more problems achieved higher scores across all outcomes, whereas requesting more hints was associated with lower conceptual, flexibility, post-math, and state test performance. Prior knowledge remained the strongest predictor in each domain (e.g., pre-conceptual $b = 0.40$, $SE = 0.03$, $p < .001$; pre-procedural $b = 0.26$, $SE = 0.04$, $p < .001$; pre-flexibility $b = 0.26$, $SE = 0.04$, $p < .001$; pre-math $b = 0.53$, $SE = 0.03$, $p < .001$; State Test 5 $b = 0.73$, $SE = 0.02$, $p < .001$).

Overall, these results clarify the mixed “replay” signal observed at the across-problem level: Immediate Replay was consistently and positively associated with students’ proximal and distal learning outcomes, while Delayed Replay was either unrelated or negatively associated when compared to Non-Replay.

\begin{table*}[t]
\centering
\caption{Regression analysis of Immediate Replay, Delayed Replay, and Non-Replay on conceptual, procedural, and flexibility scores}
\label{tab:regression_results_across_sub_replay}
\renewcommand{\arraystretch}{1.05}
\setlength{\tabcolsep}{8pt}
\small

\begin{threeparttable}
  \begin{tabular}{lll|ll|ll}
    \hline
    & \multicolumn{2}{c|}{\textbf{Conceptual}} 
    & \multicolumn{2}{c|}{\textbf{Procedural}} 
    & \multicolumn{2}{c}{\textbf{Flexibility}} \\
    \cline{2-7}
    \textbf{Predictor} 
      & $\boldsymbol{b}$ & \textbf{SE} 
      & $\boldsymbol{b}$ & \textbf{SE} 
      & $\boldsymbol{b}$ & \textbf{SE} \\
    \hline
    Intercept                          & 0.21* & 0.09 &0.2*  &0.09   &-0.03  &0.08  \\
    Immediate Replay Proportion        &1.39**  & 0.46   &1.47***  &0.41    &1.54***  &0.38  \\
    Delayed Replay Proportion          & -0.76 & 0.52    &-0.53  &0.46     &-1.1**  &0.43  \\
    Total hints requested              & -0.01* &0.00    &-0.00  & 0.00      & -0.01*** &0.00  \\
    Problem count                      & 0.01*** & 0.00   &0.01*** &0.00       &0.01***  & 0.00 \\
    Pre-conceptual Score                   & 0.4*** &0.03      &--  &--   & -- & -- \\
    Pre-procedural Score                  &--  &--       &0.26***  &0.04   & -- & -- \\
    Pre-flexibility Score                   &--  &--      &--  & --   &0.26***  &0.04  \\
    \hline
    \textbf{Model fit}  
      & $\boldsymbol{R^2}$ & \textbf{N} 
      & $\boldsymbol{R^2}$ & \textbf{N} 
      & $\boldsymbol{R^2}$ & \textbf{N} \\
      & 0.47 & 777   & 0.27 & 777   & 0.32 & 777 \\           
    \hline
  \end{tabular}
  \begin{tablenotes}[flushleft]
    \footnotesize
    \item \textit{Note.} * $p<.05$, ** $p<.01$, *** $p<.001$. All $p$-values shown are Benjamini–Hochberg FDR–adjusted. 
     Reference group: Non-replay.
  \end{tablenotes}
\end{threeparttable}

\end{table*}

\begin{table*}[t]
\centering
\caption{Regression analysis of Immediate Replay, Delayed Replay, and Non-Replay on total math and state test scores }
\label{tab:regression_results_replay}
\renewcommand{\arraystretch}{1.05}
\setlength{\tabcolsep}{10pt}
\small

\begin{threeparttable}
  \begin{tabular}{lll|ll}
    \hline
    & \multicolumn{2}{c|}{\textbf{Post-math Score}}
    & \multicolumn{2}{c}{\textbf{7th Grade State Test Score}} \\
    \cline{2-5}
    \textbf{Predictor} & $\boldsymbol{b}$ & \textbf{SE}
                       & $\boldsymbol{b}$ & \textbf{SE} \\
    \hline
    Intercept                           & -0.11   &0.18   &-0.31***    &0.04   \\
    Immediate Replay Proportion      &3.22***  &0.9  & 0.65**    & 0.22 \\
    Delayed Replay Proportion  &  -2.32*  &1.02  &-0.62*   &0.29  \\
    Total hints requested               &-0.01*   & 0.01  &-0.00**  &0.00  \\
    Problem count                       & 0.03*** &0.00    & 0.01*** &0.00  \\
    Pre-math Score                      &0.53***  & 0.03   &--       & --   \\
    5th Grade State Test Score                         &  --    & --     &0.73*** & 0.02 \\
    \hline
    \textbf{Model fit}  & $\boldsymbol{R^2}$ & \textbf{N}
                       & $\boldsymbol{R^2}$ & \textbf{N} \\

                                          & 0.55   & 777  & 0.74   &  765 \\           
    \hline
  \end{tabular}
  \begin{tablenotes}[flushleft]
    \footnotesize
    \item \textit{Note.} * $p<.05$, ** $p<.01$, *** $p<.001$. All $p$-values shown are Benjamini–Hochberg FDR–adjusted.
     Reference group: Non-replay.
  \end{tablenotes}
\end{threeparttable}

\end{table*}

\section{Discussion} 

This study used Markov chains to model students' within- and across-problem sequences, applied HMMs to uncover latent states underlying student pathways, and examined how these, as well as replay behaviors, predicted learning outcomes. We focused on both proximal (sub-constructs, total post scores) and distal (state tests) measures of learning.

For RQ1, the Markov chain analysis revealed distinct dynamics within and across problems (Figure~\ref{fig:W_P}). \emph{Within problems}, replay cycles were most prominent: students often reattempted problems immediately after solving them (e.g., \texttt{optimal}$\rightarrow$\texttt{replay\_optimal\_end} = 0.62; \texttt{sub\_optimal}$\rightarrow$\texttt{replay\_optimal\_end} = 0.34). At the same time, persistence in incomplete states (\texttt{incomplete}$\rightarrow$\texttt{incomplete} = 0.60) and replay-incomplete loops (0.47) suggest that many students repeated unproductive approaches. Some replay flows (e.g., \texttt{replay\_sub\_optimal}$\rightarrow$\texttt{replay\_incomplete}) indicate that replay could also lead back into inefficient pathways. Overall, the within-problem view highlights short-term iterations: students cycled rapidly through replays, explored alternatives, or stalled in incomplete states, with varying efficiency.

These patterns reflect trial-and-error learning, where rapid cycles test hypotheses and immediate feedback reinforces practice \cite{vanlehn2011relative}. Replay can consolidate correct procedures or expose misconceptions, but persistent incomplete states echo impasse-driven learning \cite{vanlehn1988toward}, where strategic change depends on recognizing and addressing impasses. Without reflection or feedback, replay may reinforce or otherwise signal unproductive persistence \cite{blumberg2008impasse}.

\emph{Across problems}, strategic orientations showed strong continuity (Figure~\ref{fig:A_P}). Students in incomplete states often began the next problem similarly (\texttt{incomplete}$\rightarrow$\texttt{incomplete} = 0.59), while some students repeated optimal completions (\texttt{optimal}$\rightarrow$\texttt{optimal} = 0.47). Replay behaviors also persisted across problems. These results suggest that students carry habitual problem-solving modes across tasks, consistent with prior work on ``stable modes of engagement'' \cite{chi2009active}. Cross-category shifts, though less common, showed evidence of improvement (e.g., \texttt{sub\_optimal\_end}$\rightarrow$\texttt{optimal\_end} = 0.38), supporting the idea of productive struggle \cite{granberg2016discovering}. Such transitions highlight the importance of scaffolds and feedback to help students move from inefficient to effective strategies.

For RQ2, the HMM analysis revealed four latent states: a mixed state (State 1), an incomplete-dominant state (State 2), an optimal-ending state (State 3), and a replay-dominant state (State 4). Incomplete-dominant and optimal-ending states were highly stable, with longer times within-state. Extended time in incomplete states suggests persistent challenges \cite{lodge2018understanding}, while stability in optimal-ending states indicates reuse of efficient ``problem-solving scripts'' \cite{chi1989self} and transfer of strategies across tasks \cite{winne1998studying}. In contrast, the mixed state was unstable, often transitioning into replay cycles. Replay-dominant states functioned as transitional bridges, sometimes leading to productive endings (State 3), but also back to incomplete attempts (State 2). This aligns with theories of productive failure \cite{kapur2016examining} but also could indicate behaviors associated with unproductive ``wheel spinning'' \cite{beck2013wheel}.

Replay also reflects motivational dynamics. From the perspective of Self-Determination Theory \cite{deci2012self}, replay may signal autonomy, incomplete persistence may reflect frustrated competence, and optimal-ending states may satisfy both. These findings show how latent-state modeling can distinguish stable versus transitional patterns, clarifying when persistence is productive and when scaffolding is needed.

For RQ3, regression results showed that latent states predicted both proximal and distal outcomes, controlling for prior knowledge, problems solved, and hint usage. Replay-dominant behavior (State 4) was consistently beneficial, improving conceptual understanding, flexibility, and state test performance. Optimal-ending states (State 3) also predicted higher conceptual and state test scores, suggesting that efficient completions may contribute to or indicate capacity for deeper reasoning \cite{jonsson2020gaining}. By contrast, mixed (State 1) and incomplete-dominant (State 2) states were not associated with learning gains, highlighting that persistence alone is insufficient without effective strategies \cite{o2013supporting}.

Additional covariates reinforced previously-established effects: solving more problems predicted higher scores, while greater hint use predicted lower outcomes, consistent with prior findings associating poorer performance with over-reliance on supports \cite{iannacchione2023examining}, or even the recognition that students who are struggling to learn are just more likely to seek help. Prior knowledge also remained the strongest predictor, as expected (c.f. \cite{kennedy2015predicting}).

For RQ4, Immediate Replay (retrying a problem right after solving) was strongly associated with higher performance across all outcome domains. In contrast, Delayed Replay (returning after other problems) was either weakly or negatively correlated to outcomes, particularly in relation to flexibility and state test performance. Rather than supporting transfer, Delayed Replay may reflect avoidance or gaming behaviors, particularly in cases where the behavior is observed following incomplete attempts on other problems or similar indicators of struggle. These findings extend prior work by showing that replay's benefits depend on timing and intention \cite{liu2017antecedents, harred2019long}.

\section{Limitations and Future Work} 
While our analyses captured behavioral dynamics through students’ problem-solving pathways, we did not account for the motivational factors that may influence students' strategy use \cite{su2016mathematical}. Our dataset includes pre- and post-measures of motivation as well as measures about math anxiety, which we plan to incorporate in future analyses to better understand the results of immediate versus delayed replay. That said, including these measures will offer a more comprehensive view of how cognitive (e.g., prior knowledge and outcomes), behavioral (e.g., pathways and latent states), and motivational factors jointly influence the productivity of problem-solving strategies.

\section{Conclusion} 

This study applied Markov chains and HMMs to examine students' problem-solving pathways and their relation to learning outcomes. For \textbf{RQ1}, within-problem behavior was marked by replay cycles, exploratory attempts, and incomplete loops, while across problems students tended to adopt more stable pathways (either consistently efficient completions or recurring incomplete approaches). For \textbf{RQ2}, HMMs identified four latent pathways, revealing how replay often functioned as a bridge between unstable exploration and stable completions. For \textbf{RQ3}, students who are more engaged in optimal-ending and replay-dominant pathways demonstrated stronger conceptual understanding, flexibility, and higher performance. For \textbf{RQ4}, immediate replay consistently correlated with higher performance, whereas delayed replay was either weakly or negatively associated with outcomes. These imply that flexibility may not be strengthened through all forms of replay behavior, but seems to be more correlated with more immediate engagement strategies. 

Overall, our findings suggest that flexibility develops not only through exposure to multiple strategies but also through how learners sequence and revisit their problem-solving attempts. Digital learning environments that support timely replay, encourage adaptive exploration, and scaffold transitions toward efficient strategies may be particularly effective in fostering flexibility.

\begin{acks}
We would like to thank the National Science Foundation (\#2331379), the Gates Foundation, and other anonymous philanthropies.
\end{acks}

\bibliographystyle{ACM-Reference-Format}
\bibliography{samples/reference}

\end{document}